\documentstyle[12pt,aasms4]{article}

\slugcomment{Submitted to the Astronomical Journal}
\begin{document}

\title{Escape of Lyman $\alpha$ Emission in the Starburst Galaxy Mkn 357 - 
a Wind's Far Side
\footnote{Based on observations with the NASA/ESA {\it Hubble Space Telescope}
obtained at the Space Telescope Science Institute, which is operated
by the Association of Universities for Research in Astronomy, Inc.,
under NASA contract No. NAS5-26555. These observations are
associated with program \# 8565.}}

\author{William C. Keel}
\affil{Department of Physics and Astronomy, University of Alabama, Box 870324,
Tuscaloosa, AL 35487; keel@bildad.astr.ua.edu}

\begin{abstract}
HST imaging and slitless spectroscopy are used to examine where the
strong Lyman $\alpha$ emission escapes from the interstellar medium
in the starburst galaxy Mkn 357. An H$\alpha$ image shows that the
ionized gas is mostly in a global wind, rather than associated with
the individual star-forming regions seen in the optical and UV continuum.
The Lyman $\alpha$ emission comes predominantly from the northwest side of
the wind structure spatially, and shows a significant
redshift relative to the optical lines. Both of these properties are signatures 
of seeing the line photons backscattered
from the far side of a prolate or bipolar starburst wind, fitting both with
escape calculations and evidence for winds in high-redshift galaxies
with net Lyman $\alpha$ emission. Scattering is most important
within this wind itself, rather than involving a surrounding neutral medium,
as shown by the decreasing
relative redshift of the line peak from 250 to $\approx 30$ km s$^{-1}$
between the center and edge of the detected emission. The Lyman $\alpha$
emission exhibits strong
asymmetry in comparison with both the starlight and H$\alpha$ structures.
These results add to the evidence that kinematics, rather than
gas metallicity or dust content, are the dominant effect
in determining which galaxies have strong Lyman $\alpha$ emission, and
that powerful (and perhaps episodic) starbursts are common among
Lyman-break galaxies as well as those discovered from Lyman $\alpha$
line emission.
\end{abstract}


\section{Introduction}

Lyman $\alpha$ emission plays an important role in the excitation of
gas in H II regions. Paradoxically, this role is so important that
very little Lyman $\alpha$ emission should escape to be observed.
The mean free path of a Lyman $\alpha$ photon in a typical H II region
is so small that many scatterings should occur before leaving the region
(the standard ``Case B" of photoionization calculations),
with line photons escaping either when they are emitted or scattered 
near enough to the region's boundary,
or shifted enough in frequency to increase the mean free path. This process,
resonant trapping, means that Lyman $\alpha$ photons have a much greater
probability of being absorbed by grains or converted to lower-energy
line or two-photon emission than do photons in the adjacent continuum,
so that we would expect to see virtually no line emission escaping
from star-forming regions. Such calculations were set out in detail
by Auer (1968) and Bonilha et al. (1979), showing both how effectively 
resonance-line photons are destroyed in static media, and how strongly
expansion of the medium enhances their escape. Indeed, in the static case,
Lyman $\alpha$ escape into surrounding neutral gas may serve only to 
backscatter the
line photons into the ionized region again, further increasing the odds of
their absorption by grains (Chen \& Neufeld 1994). From these considerations,
a simple model of gas in star-forming galaxies would predict emerging
Lyman $\alpha$ only from a thin outer shell plus photons far enough
into the line wings to reduce the incidence of scattering.

Empirically, the situation in star-forming galaxies proves more complicated.
Early expectations that protogalaxies could be found from intensely
strong Lyman $\alpha$ emission (Sunyaev, Tinsley, \& Meier 1978)
were dashed first by the theoretical
realization of how important scattering effects could be, and then 
by the widespread failure to find high-redshift line emitters as
powerful as simple predictions based on star-formation rate and
ionizing flux suggested. However, IUE spectra showed that some local
star-forming galaxies do exhibit significant Lyman $\alpha$ emission
(Hartmann et al. 1988, Giavalisco et al. 1996, Keel 1998). The strength
of the line
is not strongly coupled to metallicity, so something more than destruction
by dust grains in otherwise comparable systems must be taking place.
HST spectroscopy of small regions within star-forming galaxies
has shown a similar diversity, with some showing net
emission equivalent widths as large as 70 \AA\ , and some showing
no detectable emission superimposed on the stellar and interstellar-medium
absorption (Lequeux et al. 1995, Thuan \& Izotov 1997). It is especially
noteworthy that the very low-metallicity galaxies I Zw 18 and SBS 0335-052
show only net Lyman $\alpha$ absorption (Kunth et al. 1994, Thuan \& 
Izotov 1997), again showing that additional factors govern the observability
of this line.

Strong Lyman $\alpha$ emission is common from galaxies
at high redshifts, even those with no sign of an active nucleus.
In fact, Lyman $\alpha$ selection has proven to be an efficient way to find 
galaxies to high redshifts. This ranges from groups of objects at 
$z = 2.4-4.1$ (Pascarelle et al. 1996, 1998; Francis et al. 1997,
Venemans et al. 2002) to selection of galaxies at $z=4-6.5$ (Rhoads \& Malhotra 
2001, Malhotra \& Rhoads 2002, Rhoads et al. 2003, Cowie \& Hu 1998,
Hu  et al. 1999, 2003). Spectroscopy of large samples of galaxies selected from
the Lyman break shows that about 1/3 have significant Lyman $\alpha$
emission (Shapley et al. 2003). Dynamical information has shown
that some of the very large ``Lyman $\alpha$ blobs" seen at $z>2$
have properties consistent with very powerful starburst winds
(Ohyama et al. 2003).

Winds offer a promising environment for the escape of Lyman $\alpha$,
since the backscattered photons from the receding side of the wind
have the maximal wavelength shift, and therefore the largest escape probability.
This shift is enhanced when the wind scatters line emission from a 
central source
such as a concentrated starburst or AGN, so that the mean wavelength of
line photons is already redshifted before scattering in the wind.
The blueshifted side does not have this advantage, since the most
favorable scattering geometry for escape points the photons away from
the observer. Recent calculations by Ahn, Lee, \& Lee (2003) underline
this point, tracing the strongly preferred redshifting of observed
photons; under ideal geometry, multiple line peaks corresponding to
various numbers of scatterings might even be observable. Further
work (Ahn 2004) shows that widely distributed dust may be required
to fully suppress these multiple-scattering peaks, and yield the kinds of
single-peaked P Cygni profiles seen in many high-redshift systems. 
Escape of Lyman $\alpha$ from gas in winds 
fits with the wavelength asymmetry often seen
in spectra of both nearby star-forming regions and high-redshift
star-forming galaxies. As stressed by Mas-Hesse et al. (2003) 
from analysis of HST
aperture spectra, the dynamics of surrounding {\it neutral} gas may be
much more important in the escape of Lyman $\alpha$ than the
galaxy's dust content, and this fact could shape our selection of
high-redshift galaxies from line emission strongly in favor of
objects with global winds. This recognition further complicates
the situation in star-forming galaxies; global starburst winds
give ample amounts of outflowing gas, but these winds can be so hot
as to have a very small and patchy neutral component.

Understanding where and how Lyman $\alpha$ emission escapes from star-forming
systems is thus important for our interpretation of the early history
of galaxies, as well as for the physics of the interstellar medium. 
However, spatially resolved measurements of Lyman $\alpha$ 
from nearby galaxies present formidable challenges. Geocoronal Lyman $\alpha$ 
emission compromises detections at very low redshifts, while foreground absorption
from the Milky Way's own H I can affect measurements of galaxies
with redshifts up to several hundred km s$^{-1}$ along many
lines of sight. The lack of suitable
narrowband filters has made spectroscopic approaches the most
effective way to seek line emission. A handful of objects were observed
using IUE, as noted above, and further spectroscopy of compact
galaxies and individual star-forming regions using the instruments
on board HST has given detections and line profiles in several instances
(Kunth et al. 1998, Lequeux et al. 1995, Mas-Hesse et al. 2003). 
In each case, the emission shows a P Cygni profile, even when
stellar background light is not an issue, implicating kinematic processes
in what we can see of Lyman $\alpha$.

A scanning observation of M33 by {\it Voyager 2} (Keel 1998), with
one dimension of spatial information,
showed diffuse Lyman $\alpha$ emission with no peak at the
location of the very luminous H II region NGC 604, leading to the
conjecture that the line arises in the diffuse interstellar medium
rather than from individual H II regions. More recently, Mas-Hesse
et al. (2003) have used HST spectroscopy to trace the extent of
P Cygni structures and wavelenght shifts in star-forming galaxies.
The new UV imaging capabilities brought by HST's Advanced Camera for
Surveys (ACS) have been exploited by Kunth et al. (2004) to attempt
imaging of Lyman $\alpha$ emission in nearby star-forming systems,
an approach complemented by the spectroscopic technique used here.
This paper describes the use of slitless Lyman $\alpha$ spectroscopy,
with acompanying images in the continuum and H$\alpha$, to examine the
spatial and velocity structure of Lyman $\alpha$ emission from
a luminous starburst galaxy. Strong radiative-transfer signatures are seen,
and the line emission comes from the far side of a 
starburst wind, probably a global superwind. This fits with 
less direct evidence from
high-redshift galaxies, and suggests that such winds were commonplace
in the early Universe.

\section{Observations}

\subsection{Mkn 357 as a Lyman $\alpha$ Emitter}

Mkn 357 (PG 0119+229) is a promising system in which to study the escape
of Lyman $\alpha$ emission from star-forming galaxies. Its optical
emission lines show the line ratios of a starburst, at redshift $z=0.053$
and thus well beyond contamination of its emission by 
Lyman $\alpha$ absorption from gas in the Milky Way. Ground-based
imagery shows a close companion and extensive, asymmetric H$\alpha$ emission.
An archival IUE observation (Keel 1998) shows very strong 
Lyman $\alpha$
emission, guaranteeing enough signal to trace its spatial as well
as velocity structure using HST. As one of the UV-brightest starbursts
at such redshifts, Mkn 357 has also been a high-priority FUSE
target to seek emerging Lyman-continuum radiation.  

I present here a four-way test for the roles of resonant trapping
and velocity structure in the escape of Lyman $\alpha$. A slitless
spectrum around Lyman $\alpha$ provides both the spatial and spectral
structure of escaping line photons. An H$\alpha$ image shows the distribution 
of ionized gas, to distinguish the roles of radiative transfer
and structure in the gas.
The impact of dust extinction can be assessed from comparison 
of broad-band UV
and red-light images. These continuum images also show where potential
ionizing clusters are, and thus distinguish roles for an outflow wind
and discrete H II regions in the ionized gas. A ground-based spectrum
covering H$\alpha$ sets the redshift zero point in assessing the
velocity dependence of the escape of Lyman $\alpha$.
 
Using H$_0$=70 km s$^{-1}$ Mpc$^{-1}$ and a flat Universe with $\Omega_M=0.3$,
Mkn 357 is seen at a luminosity distance of 237 Mpc (distance modulus
36.87), and the projected linear scale is 1.03 kpc arcsec$^{-1}$.
This gives linear pixel scales of about 50 pc for the WFPC2
images (on the PC CCD) and half that for the STIS ultraviolet data.
With this distance, Mkn 357 is fairly luminous for such a compact 
system, with $M_B=-21.43$, $M_V = -21.67$. This makes it about
twice as luminous as the brightest system observed by Mas-Hesse
et al. (2003), namely IRAS 0833+6517 at $M_B=-20.8$, which shows 
closely comparable metallicity.

\subsection{Lyman $\alpha$ spectroscopy}

Lyman $\alpha$ was observed using the ``slitless" spectroscopic mode of
STIS on board HST. In practice, the wide 2" slit (aperture 52X2) was in place,
to exclude geocoronal Lyman $\alpha$ and O I $\lambda 1304$ emission
while including virtually all of the Lyman $\alpha$ emission
from Mkn 357, which is observed at 1280 \AA\ . The telescope orientation
put the slit in position angle $-42^\circ$, roughly along the major
axis of Mkn 357 and crossing some of the companion as well.
The FUV MAMA 
detector and G140M grating gave a spectral scale of 0.0529
\AA\ pixel$^{-1}$, with the spectral resolution set by the object's structure
parallel to the dispersion. The sampling in the purely
spatial direction was 0.029"/pixel.
The spectral range was 1245-1299 \AA\ .
The exposure was 1560 seconds, cut short of the planned 4133 seconds 
by an electronics reset. The total Lyman $\alpha$ flux from the
archival IUE observations (SWP 18199 and 45198) was used as a consistency 
check on the STIS flux scale given this timing issue, using the pipeline
conversion factor {\it diff2pt} = 0.084 to derive a total in-slit flux
from the values calibrated for surface brightness.
The STIS total for a 200-pixel swath along the slit is $9.4 \times
10^{-14}$ erg cm$^{-2}$ s$^{-1}$, somewhat less than the IUE value of
$1.6 \pm 0.3 \times 10^{-13}$, but not by a large enough factor to
require a calibration difference, since some additional Lyman $\alpha$
might fall beyond the 2" STIS wide slit.
  
\subsection{HST Imaging}

A far-UV image was also obtained with the STIS FUV MAMA, to complement
the Lyman $\alpha$ data by tracing the morphology seen in hot stars,
and particularly the presence and locations of massive ionizing clusters. 
The SrF$_2$ filter
gave a passband about 120 \AA\  wide centered near 1460 \AA\ , with
no significant red leak.
The total exposure was 2160 seconds, with spatial sampling of 0.0247 arcseconds
pixel$^{-1}$. Both the STIS and WFPC2 observations for this program were obtained 
on 12 September 2000.

The WFPC2 linear ramp filter (LRF) FR680N was used to isolate H$\alpha$
emission, with a passband spanning 89 \AA\  FWHM. The central wavelength
was set to 6922 \AA\ , 10 \AA\  redward of the H$\alpha$ peak wavelength
found spectroscopically. This offset, well within the FWHM of the passband
and retaining about 90\% of the peak transmission, allowed Mkn 357
to be observed on the PC CCD, with nearly Nyquist sampling of the image
at this wavelength. A total of 1577 seconds' exposure was obtained
in two images (with the second terminated by a guiding ``loss of lock" shortly
before its scheduled end), which were combined for cosmic-ray rejection.

The continuum in red light was observed in a single WFPC2 exposure through the
F702W filter, lasting 500 seconds. The pivot wavelength for this passband is 
6917 \AA\ , close enough to redshifted H$\alpha$ to make first-order
color terms in the 
continuum subtraction irrelevant. Cosmic rays were removed by interactive
editing, guided by the near-Nyquist sampling of the PC PSF at this wavelength,
and the implausibility of bright objects which do not appear at all in the 
narrowband image or UV STIS data. The galaxy was approximately centered in 
the PC CCD for this image, again taking advantage of its better
sampling on a bright target. This pointing included the companion (or tail)
of Mkn 357, which fell off the chip in the H$\alpha$
images. Alignment of the broad- and narrow-band images at the
same telescope orientation should be a pure translation except for
effects of optical distortion. Indeed, an image shift matches the
structures in common as well as the difference image will show, without
incurring the additional
loss in resolution in resampling to correct for the small
differential distortion between the two image locations. The continuum
image was scaled for subtraction using the integrated H$\alpha$ and [N II]
equivalent widths from the INT spectrum (section 2.4), 
weighted by the LRF passband shape. To the approximation that all
line emission within the F702W passband is included in the LRF
filter, the two exposures contain enough information to separate
the two contributions by linear combinations in which the constants are
set by the relative filter widths and exposure times.

The resulting net H$\alpha$ image shows ionized gas in a structure strikingly
different from the starlight (section 2.2). This image was in turn 
subtracted from the scaled $R$ image to give a nearly pure continuum 
distribution (except for the weaker [O I] and [S II] lines)
for more accurate comparison with the UV starlight.
 
The red-light image was aligned with the STIS UV continuum data using
9 compact sources (star clusters) which are bright in both passbands,
with a fit accuracy per source of about 0.7 STIS pixel in each
coordinate. This registration allows
accurate ratio maps and photometry of the clusters, which were measured 
within circular apertures of radius 0.1" using local background subtraction
out to 0.25" radius. Their magnitudes were set to roughly the Johnson
$R$ zero point for the WFPC2 F702W data, and in the ultraviolet using the 
magnitude convention based on $m = -2.5 \log {\rm F}_\lambda -21.1$
where F$_\lambda$ is measured in ergs cm$^{-2}$ s$^{-1}$ \AA\ $^{-1}$.

\subsection{Ground-based spectroscopy}

To define the systemic redshift accurately, and derive line ratios sensitive
to the roles of stellar photionization and shock heating, I use a long-slit
spectrum obtained with the 2.5-m Isaac Newton Telescope on La Palma, from
December 1985. The 1.0" slit was oriented in position angle 135$^\circ$,
to include the companion object in the 1800-second exposure. The 
Intermediate-Dispersion Spectrograph was used with grating ruled at 
831 lines mm$^{-1}$, delivering a spatial sampling of 0.65" per pixel and
spectral sampling of
1.1 \AA\ per pixel at 2.2 \AA\  FWHM resolution on a GEC CCD. The spectral 
range covered
6642-7289 \AA\, just including the shock-sensitive [O I] $\lambda 6300$
line observed at 6648 \AA\ . The systemic emission-line redshift
from H$\alpha$ is $z=0.0531 \pm 0.0001$, with the uncertainty derived from
changes when different assumptions about the relative widths of H$\alpha$
and the partially overlapping [N II] lines are used. The spectrum also
shows a small net velocity gradient, 
of observed amplitude 53 km s$^{-1}$ across
the 18" span of detected emission, with the northwestern side receding. No 
line emission from the companion to the southeast was detected in this spectrum.

This spectrum gives integrated equivalent widths (in the observed frame)
of 242 \AA\  for H$\alpha$ and 30 \AA\  for [N II]$\lambda 6583$.
The profile of
H$\alpha$ is best represented as the sum of two components with Gaussian
FWHM of 3.3 and 10.2 \AA\  (140 and 442 km s$^{-1}$), with intensity
ratio 1.9:1. In light of the WFPC2 H$\alpha$ imagery, the broader
component might reflect emission from escaping wind material. However,
shock ionization in such an outflow is not important in the integrated line
ratios, with [O I]/H$\alpha$=0.02 and [S II]/H$\alpha$=0.16, neither of which is
an unusual value for H II regions or young starbursts. This suggests that
line emission from a wind is dominated either by {\it in situ}
photoionization or line emission scattered from dust in the wind.

The optical emission lines allow an estimate of the current gas-phase
metallicity, using the traditional bright-line approach (Pagel et al. 1979,
Edmunds \& Pagel 1984, McCall et al. 1985).
The INT spectrum was supplemented for this purpose by the blue portion 
of the Kitt Peak IIDS spectrum reproduced by Keel (1985).
Using the relations from Pilyugin (2000, 2001) gives 12 + log (O/H) = 8.35,
roughly half solar with the ``new" solar calibration. This is in the
range seen nearby in such small starburst systems as NGC 1569 (e.g., Devost,
Roy, \& Drissen 1997).

\section{Starburst Properties}

\subsection{Morphology and Interaction}

Ground-based imaging suggested that Mkn 357 was an inclined disk system
with tidal tail and companion. These new data show a much more irregular
structure, clearly not a normal spiral galaxy. The companion, about
12" away in projection (12.3 kpc) does not show a central concentration,
having a few diffuse bright knots within an amorphous glow that blends
into the bridge connecting it to Mkn 357 itself. Two potential star
clusters appear within the tidal debris around this bridge. These features
are shown in Fig. 1, comparing UV and red images. 
Mkn 357 is not a very large galaxy; half
the $R$-band light is emitted within a projected radius of 0.9"
(0.94 kpc), while the WFPC2 $R$ image detects continuum light out to 
only a 6.5-kpc radius along the most extensive directions. Mkn 357 falls in 
the range of size and luminosity often associated with blue compact
dwarf galaxies.

Within Mkn 357, both the UV and $R$ images reveal a rich population of luminous
star clusters, concentrated to the center, further evidence of its
starburst nature. Local starburst systems are often taken as analogs of
high-redshift galaxies. This is probably useful as regards high levels of
energy input from star-forming regions, but will not generally be
accurate in detail, since even rather quiescent galaxies today incorporate
the heritage of a Hubble time's worth of gradual ISM enrichment and,
quite likely, episodic infall of less-processed material.

Further detail in the structure and history of Mkn 357 may be gleaned from
a simple tomographic approach, suggested by the work of Katz-Stone \& Rudnick
(1997) on radio maps and foreshadowed by 
Zwicky's (1955) photographic decomposition of galaxy images. In a simple system 
consisting of two populations
which differ in both color and spatial structure, an empirical
separation may be performed by subtracting variously scaled versions
of an image in one band from an image at a very different wavelength,
with the purest view of one population's structure found at the
largest scaling factor for which significant oversubtraction does not
occur. In realistic situations, a single scaling factor may be inadequate
for a whole object, but useful insights can be gained from even approximate
decomposition. For Mkn 357, the UV and R images decompose largely into
two components differing in both structure and color (Fig. 2).

The bluer component includes many of the prominent star clusters
(section 3.3), and
shows a rounder outer shape, with typical axial ratio 0.9. Existing ultraviolet
observations of prominent starbursts 
(Marcum et al. 2001, Hoopes et al. 2004) suggest that this
component may include scattered light from grains, perhaps including
grains mixed with a starburst wind. The strong forward-scattering phase
function of typical grains in the UV will weight this scattering
toward those in the foreground of the stars (Witt \& Gordon 2000), 
so this distribution
is not necessarily representative of the overall grain distribution.
The redder component is much more
elongated, at least in projection, with typical axial ratio 0.3. Notably, 
there are several star clusters
which appear only in the redder population, particularly at the southeastern
end. The brightest of these might represent the nucleus of a pre-existing
system which has seen strong tidal disruption, or be a remnant of a
previous burst of star formation. This cluster is in fact redder than
the rest of this component, leaving a hole in the blue component 
in a decomposition which works well almost everywhere else.

\subsection{H$\alpha$ Morphology: a Starburst Wind}

The continuum-subtracted H$\alpha$+[N II] image (Fig. 3) shows structure
quite different from that seen in the continuum. The dominant feature
is an irregular spray of emission, less flattened in projection than the
continuum light but more elongated than the blue tomographic
component, centered near the brightest star clusters. The brightest
emission traces a U shape, with peaks close to (but not centered on)
several bright
clusters. This structure resembles the H$\alpha$ appearance seen on a slightly
smaller scale in
NGC 1569 (Devost, Roy, \& Drissen 1997; Martin 1998).
However, there is not a detailed
match of continuum and emission-line features; in particular, the image
does not show distinct H II regions associated with most of the UV-bright
clusters. This is a common situation in galaxies with strong and
ongoing star formation. Numerous similar examples appear in the 
collection by Marlowe et al. (1997) and the extensive atlas
by Gil de Paz, Madore, \& Pevunova (2003).

The prominent dark lane in H$\alpha$ does not seem to be associated with
extinction, as traced either by a lack of star-forming knots or the UV/R
broadband color. This again is reminiscent of NGC 1569,
where a similar open structure appears in both optical line emission and
the radio continuum, so it must reflect a genuine lack of ionized gas.
In that system, this can be attributed to wind sweeping (Martin, Kobulnicky, 
\& Heckman 2002),
a picture which also makes sense in Mkn 357 given the other evidence
for significant outflow.

These characteristics suggest that much of
the optical line emission arises in a global
starburst (super)wind rather than from individual H II regions.
This wind may be reflected in the broad (400 km s$^{-1}$ FWHM) component
seen in the integrated H$\alpha$ profiles (section 2.4). Such an outflow
may be sampled (in part) by the velocity gradient in H$\alpha$, which may
have a counterpart component in Lyman $\alpha$ as well (section 4.2).
In some starbursts with powerful outflows, the bipolar structure is
evident only with detailed kinematics (e.g., Heckman et al. 1995), 
and in Mkn 357 the combined Lyman $\alpha$ asymmetries in intensity and
velocity may offer the best evidence for a bipolar outflow geometry.
Comparison with kinematic and morphological features seen in the gas
of star-forming galaxies on a wide range of scales (Legrand et al. 1997,
Gil de Paz et al. 1999, Silich et al. 2002, L{\'{\i}}pari et al. 2004)
points up an ambiguity in the scale of outflows - is there really
a single global superwind, or an amalgamation of expanding, overlapping 
shells around individual star-forming regions? These shells or bubbles
would not necessarily escape the galaxy potential (as a wind, properly
called, would), and would not individually extent over a large part of the
galaxy image. Definitive discrimination between local and global outflows
would require high-resolution kinematic data or X-ray observations. Weak
evidence suggesting a global pattern may come from the degree or
redshifting of the peak Lyman $\alpha $ emission and its systematic
change with distance from the galaxy center.

\subsection{Clusters in the Starburst}

Twenty bright clusters are well enough separated in these images for
individual aperture photometry. These span a wide range in both
brightness and color, with magnitudes $R = 19.0-26.4$, $m(1400) = 19.2-26.0$.
The color range is not connected to magnitude, and spans 
$m(1400) - R = -2.1$ to +4.6. One luminous red cluster is much larger
than the others and might represent the core of a former
interacting component. Before any internal reddening correction,
the brightest clusters reach $M_R=-18, M_{1400}=-17.6$. This is already
luminous enough to suggest that the clusters must be very young, since
such systems as NGC 4038/9 have clusters only up to about absolute 
magnitude -14 in these bands (Whitmore et al. 1999).

The combination of young clusters with a lack of discrete H II regions
suggests that this system is in a stage comparable to the late-stage
evolution of individual H II regions discussed by Bosch et al. (2002),
when stellar mechanical-energy input dominates the gas dynamics.

\section{Lyman $\alpha$ Emission}

\subsection{Spatial distribution of Lyman $\alpha$}

The distribution of Lyman $\alpha$ emission is well traced in the
direction along the STIS slit, and shows a striking asymmetry
and structural differences from both H$\alpha$ and the starlight.
Fig. 4 compares traces along the slit in these features and in the
composite quantity of emission equivalent width. The emission-line
flux was evaluated over a 2.1-\AA\  band, with continuum taken from a
similar band just redward of the line location. This was intended to
span the entire observed velocity range of line emission, but some
variation in the absorption component may be included.

The Lyman $\alpha$ emission distribution is very asymmetric, strongest
on the northwestern side (which is receding according to
the H$\alpha$ and [N II] velocity gradient). The northwestern side of the
emission encompasses 72\% of the total, and emission can be detected
three times as far from the core on this side (where the line
trace disappears into the noise about 8" from the core). The asymmetry
is similar to that observed in Haro 2 by Mas-Hesse et al. (2003).

There is a narrow spike of line emission at the starlight core, but
the rest of the Lyman $\alpha$ structure lacks continuum counterparts.
Broadly, this means that it does not come from discrete H II regions
around the UV-bright clusters, but from a more diffuse 
distribution of gas. Perhaps more interesting, structures traced
by Lyman $\alpha$ and H$\alpha$ emission do not coincide. Quite generally,
this means that radiative transfer (likely modulated by the overall
velocity field) dominates over either the distribution of ionizing
sources or the gas distribution in determining where Lyman $\alpha$
escapes from the galaxy. This remains true even if nonradiative
ionization (for example, by shocks within or at the interface of the
wind) is important, since it is the difference in structure between
these two recombination lines that matters.

The Lyman $\alpha$/H$\alpha$ flux ratio ranges from a low value of
0.56 (in a broad region 0.3--3" southeast of the core) to a high
of 21, seen on the edge of the emission region 1.4--3.1" northwest.
Redistribution of the line photons in wavelength during scattering
is so effective that this
line ratio can locally somewhat exceed its pure-recombination value.

\subsection{Lyman $\alpha$ velocity structure - scattering in a wind}

The Lyman $\alpha$ line shows prominent absorption as well as emission wherever
the signal-to-noise ratio is sufficient, with a characteristic 
P Cygni profile. Much of the absorption must be due to radiative-transfer
effects within the gas itself rather than to superposition on
stellar photospheric
absorption or absorption of background starlight. 
This is shown by the large velocity offset of the emission
from the absorption core, the spatial persistence of the absorption 
well beyond the compact 
UV continuum distribution, and its change with distance from the center
and thus with column density of scattering ions. The effects of this
absorption on the line profile are difficult to extract without
more complete information, since it may combine three effects -
absorption of the continuum on stellar photospheres, absorption
by foreground H I in the outskirts of Mkn 357, and an attenuation
feature of continuum photons due to the large scattering cross section
near the line center (as described by Chen \& Neufeld 1994). 
For this reason, I concentrate here on properties extracted
directly from the spectrum. These are the results which are
both most robust, and most relevant to the comparison with
high-redshift galaxies.

To extract peak wavelengths from the asymmetric emission line in such
noisy data, a high-order polynomial was fit to the data points from
1275--1285 \AA\ ; orders from 21-45 gave very similar results, with
the values for 35 being quoted. The peak wavelength was taken as the
peak of this fit (noting that multiple peaks are seen at some slit locations).
This smoothed fit incorporates all the statistically significant
structure in the emission-line profile, as can be seen visually
from examination of the residual pattern upon subtracting the fit
from the original data (Fig. 5), which accords with the Poisson
errors for the STIS data delivered by the calibration pipeline.

The peak wavelength was measured by following the maximum value of the
polynomial fits from one spatial increment to the next (using the IRAF
routine {\tt reidentify}). Errors for this peak estimate
were assessed in a Monte Carlo manner, from 1000 realizations of appropriate
Poisson errors added to the smooth fit for each line of the spectrum.
The dominant velocity feature is
a redshift of the peak velocity, greatest in the middle of the
detected line emission rather than near the galaxy core. As shown in Fig. 6, 
the offset of the Lyman $\alpha$ peak from systemic velocity is as
large as 250 km s$^{-1}$ near the middle of the emission region, 
declining to
$\approx 30$ km s$^{-1}$ toward the edges of the detectable emission.
In these regions, the emission also shows evidence for multiple
velocity components, so genuine kinematic structure may be playing
a role in the inferred wind. 
As noted above, the Lyman $\alpha$
emission region appears very asymmetric with respect
to both the galaxy's starlight and the recombination emission seen
at H$\alpha$, which indicates that radiative-transfer effects operate
in a very asymmetric way. As noted above, if most of the line
photons originate in a small core region before scattering, the
typical line Doppler shift along our line of sight will be twice
the characteristic expansion velocity, due to the backscattering
``moving mirror" geometry.
	
This analysis has treated the STIS spectrum as a pure wavelength-slit position
measure. The wide slit means that location of structures parallel to the 
dispersion may play a role. Such a role is unlikely to account for much
of the structure in peak wavelength, simply because it would require
a large U-shaped emission region with no counterpart at other wavelengths.
There is very likely a spatial component to the apparent line widths,
whose extent depends on the angular span of Lyman $\alpha$
emission. Much of the width of the Lyman $\alpha$ peak could
arise in this way if the emission and UV continuum span comparable angles. 
However, the sense of the Lyman $\alpha$ velocity structure
runs opposite to the direction that would result from resolved
galaxy structure if the emission roughly follows the UV continuum light,
suggesting that the velocity {\it structure} is unlikely to be
compromised. One test of the extent of this smearing by spatial structure
considers the widths of interstellar absorption features seen just
blueward of Lyman $\alpha$, of which the strongest unblended one in this
spectral range is Si III $\lambda 1206.5$. At the nucleus, this
line shows a FWHM of 2.5 \AA\ (or 
590 km s$^{-1}$), and is slightly deeper in core attenuation and stronger 
overall than seen in Haro 2 (Mas-Hesse et al. 2003). This line shows
a core-to-edge change in radial velocity no greater than 60
km s$^{-1}$ across 65 STIS pixels (1.7 arcseconds), a span limited 
by the background starlight. Since the Lyman $\alpha$ centroid
changes by nearly 300 km s$^{-1}$ across the same locations,
confusion by spatial structure of the galaxy is unlikely to be important.

The sense of the velocity offset across the galaxy core is the same as that
of the H$\alpha$ velocity gradient, although the span and slope are 
much greater.
The ground-based spectrum shows only a 50 km s$^{-1}$ range in H$\alpha$
velocity across the entire system.

\section{Discussion}

A comparison among Lyman $\alpha$, H$\alpha$, and multicolor continuum
data for the luminous and metal-rich starburst system Mkn 357
shows that the Lyman $\alpha$ emission is associated with a region
which is very asymmetric about the galaxy center and likely associated
with the receding half of a global wind. The remains some ambiguity 
about whether this might be overlappig winds from individual
star-forming events, or a global wid which is no longer bound in the
galaxy potential. Some evidence in favor of a global superwind comes
from fairly cohesive velocity structure in Lyman $\alpha$ emission
across the galaxy, suggesting one mass of gas is responsible.

These conclusions fit with the results of several previous studies (section 1) 
in showing that
strong Lyman $\alpha$ emission is more a sign of kinematics
(favoring outflow geometries) than metallicity (through the abundance of
grains). Applied at high redshifts, this means that starburst outflows
are very common among Lyman-break galaxies as well as the narrow-line
Lyman $\alpha$ emitters which are common at $z>2$.
They also fit with the redshift behavior seen from various
spectral features in Lyman-break galaxies
(Franx et al. 1997, Adelberger et al. 2003
Steidel et al. 1998, 2003).
As Steidel et al. (2003) note, in the presence of a strong outflow,
interstellar absorption lines will be preferentially blueshifted
compared to the systemic value, while backscattered Lyman $\alpha$
is preferentially redshifted. Since the strongest absorption features
in the easily-sampled mid-UV range can have significant interstellar
components, the systemic redshift can remain uncertain by several
hundred km s$^{-1}$.

These results also reiterate that Lyman $\alpha$ cannot be used as
a simple tracer of the star-formation rate, reinforcing the arguments
of Kunth et al. (2004). They find knots and diffuse structures not
obviously related to the SFR distribution, fitting in some cases with
a wind picture. From their sample, ESO 338-IG04 in particular shows
structures much like the wind in Mkn 357. Comparison with these
more local objects points up the role of neutral outflowing gas, 
which may be only a minority constituent of hot, mostly ionized
winds from objects undergoing powerful starbursts, and adds further
complexity to the interpretation of Lyman $\alpha$ emission from
high-redshift objects. Not only active star formation and outflows,
but the reservoir of additional gas or multiphase nature of the
outflow, may play significant roles.

\acknowledgments
This research was supported by NASA through STSCI grant GO-8565.01-A.
The Isaac Newton Telescope is operated
by the Royal Greenwich Observatory on behalf of the SERC at the Spanish
Observatorio del Roque de los Muchachos. Patrick Treuthardt provided
useful assistance in evaluating the spectral errors and searching
the literature. I thank Rogier Windhorst for a critical reading of the
manuscript. The referee, D. Kunth, also contributed to the clarity and
completeness of the discussion.

\clearpage

\clearpage
\figcaption
{Montage of ultraviolet and red-light images of Mkn 357 and its
companion structure, rebinned to the same scale and orientation. The field
show is $15.4 \times 17.2$ arcseconds; north is $6^\circ$ clockwise from the top.
A logarithmic intensity display was used to moderate the contrast;
even so, the inner structure is best shown in the lower panels,
enlarged by a factor 4 and stretched for a wider intensity range.
\label{fig1}}

\figcaption
{Tomographic decomposition of Mkn 357, from the UV and red-light images
after rebinning to the same scale and comparable resolution. The redder
component is much more elongated than the UV-bright blue component of the
galaxy, which may include scattered light from grains seen at a small
angle in front of the bright clusters. For clarity, the intensity
mapping in each case is logarithmic with a small offset from zero.
The field is $6.0 \times 4.9$", 
and north is $6^\circ$ clockwise from the top.\label{fig2}}

\figcaption
{The H$\alpha$+[N II] structure of Mkn 357. Left, the net emission-line
image after continuum subtraction based on the $R$-band F702W image,
displayed with an offset logarithmic scale. Right, the same image
shown as contours overlaid on the red-light continuum image, corrected for
emission-line contamination, to show the different structures in starlight
and ionized gas. The continuum image has likewise been displayed with
a logarithmic intensity scale. As in Fig. 2, the field is $6.0 \times 4.9$", 
and north is $6^\circ$ clockwise from the top.
\label{fig3}}

\figcaption
{Comparative distributions of Lyman $\alpha$, H$\alpha$ and the ultraviolet
continuum along the slit. Upper left, trace of Lyman $\alpha$ emission
intensity (in arbitrary units) as observed and with 0.5" boxcar
smoothing. The line flux was summed over a 2.1-\AA\  band, with continuum
subtracted based on an equal-width band just to the red. Upper right,
the Lyman $\alpha$ slice is overlaid with the ultraviolet continuum,
both from the spectrum near 1300 \AA\  and from the STIS direct image.
The two contionuum traces were used to align the data sets, and their
agreement in structure confirms that the positioning of the STIS and
WFPC2 data is reliable. There is a Lyman $\alpha$ emission spike near the 
starlight core, but the Lyman $\alpha$ 
profile is much more extended and shows structure on both sides of tis
core not related to UV starlight
Lower left, the Lyman $\alpha$ and H$\alpha$ profiles are compared.
Again, the detailed correspondence is poor, with peaks in the two lines
not aligned. The increasing Lyman $\alpha$/H$\alpha$ ratio with
distance from the center is a sign of radiative-transfer effects.
Lower right, the trace of Lyman $\alpha$ mission equivalent width,
in the emitted frame, is shown with and without 0.5" smoothing.
The increase in equivalent width going outwards indicates some
combination of radiative transfer and ionization by non-local sources.
\label{fig4}}

\figcaption
{Lyman $\alpha$ profile from the STIS wide-slit observation, displayed with
linear intensity scales after $2 \times2$-pixel block averaging. The
data are shown as obtained and in the polynomial fit used to derive
peak wavelength, along with the difference between them (amplified
by a factor 10 in amplitude for display). The region shown spans
10.07 \AA\  on the horizontal axis and 7.5" in the spatial (vertical)
direction, oriented so that the northwest end of the slit (in position
angle $-42^\circ$) is at the top. The absorption component of Lyman
$\alpha$ is prominent left of center in each panel.
\label{fig5}}

\figcaption
{Lyman $\alpha$ profiles along the slit. Each is centered at the indicated
distance from the UV continuum peak, in arcseconds, with negative values
lying to the southeast along position angle $138^\circ$, and represents data
summed over the indicated slit length in arcseconds. The heavy line shows
the polynomial fit used to derive peak wavelengths, and the difference
between data and fit is shown at the bottom. For comparison, all spectra
have been normalized to the same peak intensity. The inner regions show
pronounced P Cygni profiles, with the peak velocity redshifted by $\approx 250$
km s$^{-1}$ in a velocity scale set from the integrated H$\alpha$ and [N II]
emission. Far from the core, the emission has multiple statistically-significant
components, including both redshifted and near-systemic values; no
blueshifted components appear.
\label{fig6}}

\figcaption
{Velocity peak of Lyman $\alpha$, as defined by the fits in Fig. 5, 
traced along the slit track. Error analysis is complicated by the
profile fitting, but the envelope of the velocity error may
be seen from local excursions as the intensity drops. Adjacent points are 
averages of two
original lines of the spectrum, and are almost completely independent
statistically (except for a small crosstalk due to rebinning in accounting
for distortion along the slit). The shaded curve at the bottom indicates the
net intensity of Lyman $\alpha$ emission at each location, to show the
alignment with emission structures and the relative importance of
different velocity-offset regimes. The zero point was set by the 
mean redshift of H$\alpha$+[N II] emission, and should be accurate
to $\pm 30$ km s$^{-1}$. The error envelope indicates $\pm 1 \sigma$
for the main peak, and was evaluated with a Monte Carlo approach
based on the pixel-to-pixel noise, adding realizations of this
noise to the smoothed fits and remeasuring the peak. The error depends
on not only the line flux, but noth the width and profile of the
emission.
\label{fig7}}

\end{document}